\documentclass[11pt]{article}

\newcommand{\BABARPubYear}    {04}

\newcommand{\BABARProcNumber} {031}
\newcommand{\SLACPubNumber} {10677}
\newcommand{\LANLNumber} {0409006}

\newcommand{\ra}{\rightarrow}
\newcommand{\prl}{Phys.Rev.Lett. }
\newcommand{\prd}{Phys.Rev. D}

\def\kppim{\mbox{$K^+ \pi^-$}}
\def\kppiz{\mbox{$K^+ \pi^0$}}
\def\kzpip{\mbox{$K^0 \pi^+$}}
\def\kzpiz{\mbox{$K^0 \pi^0$}}
\def\pippim{\mbox{$\pi^+ \pi^-$}}
\def\pippiz{\mbox{$\pi^+ \pi^0$}}
\def\pizpiz{\mbox{$\pi^0 \pi^0$}}
\def\kpkm{\mbox{$K^+ K^-$}}
\def\kpkzb{\mbox{$K^+ \overline{K}{}^0$}}
\def\kzkzb{\mbox{$K^0 \overline{K}{}^0$}}

\input pubboard/babarsym

\long\def\inst#1{\par\nobreak\kern 4pt\nobreak
    {\it #1}\par\vskip 10pt plus 3pt minus 3pt}

\begin{document}
{\pagestyle{empty}

\begin{flushright}
SLAC-PUB-\SLACPubNumber \\
\babar-PROC-\BABARPubYear/\BABARProcNumber \\
hep-ex/\LANLNumber \\
August, 2004 \\
\end{flushright}

\par\vskip 4cm

\begin{center}
\Large \bf Rare Hadronic B Decays
\end{center}
\bigskip

\begin{center}
\large 
Liliana Teodorescu\\
Brunel University, West London \\
Uxbridge, UB8 3PH, United Kingdom \\
(from the \lbabar\ Collaboration)
\end{center}
\bigskip \bigskip

\begin{center}
\large \bf Abstract
\end{center}
A review of recent results  on branching 
fractions  and time-integrated CP asymmetries of rare $B_u$ and $B_d$ decays 
to mesonic final states 
from \babar\, Belle and CLEO experiments is presented.

\vfill
\begin{center}
Contributed to the Proceedings of the 6$^{th}$ International 
Conference on Hyperons, Charm and Beauty Hadrons \\
6/27/2004---7/03/2004, Chicago, USA
\end{center}

\vspace{1.0cm}
\begin{center}
{\em Stanford Linear Accelerator Center, Stanford University, 
Stanford, CA 94309} \\ \vspace{0.1cm}\hrule\vspace{0.1cm}
Work supported in part by Department of Energy contract DE-AC03-76SF00515.
\end{center}

\section{Introduction}

\hspace{0.5cm}The term ``rare'' B decays  refers to B decays with 
small branching fractions (BF), usually less than $10^{-5}$. These are 
B decays that do not involve $b\rightarrow c$ transitions that are
favoured by a large Cabibbo-Kobayashi-Maskawa (CKM) matrix element. 
They  proceed, typically, through tree $b\rightarrow u$ transitions that
are suppressed by a small CKM matrix element and/or through second order (penguin) 
$b\rightarrow s$ or $b\rightarrow d$ transitions. When both tree and
penguin transitions
are present, the interference between their amplitudes might be significant, 
leading to direct CP violation.

The direct CP violation in $B$ decays occurs when the decay rate of a $B$ into a 
final state is different than the decay rate of its antiparticle in the 
charge-conjugate final state. The difference is quantified by the 
time-integrated CP asymmetry ($A_{CP}$) defined 
as the difference of the two decay rates normalised to their sum.

\vspace{0.2cm}
Rare $B$ decays into hadronic final states are of interest for both Standard Model 
and new physics studies.

Precise tests of the Standard Model predictions, including tests of the CP violation 
mechanism, can be achieved with these small amplitude processes.  Many of these
processes are 
potential tools for extracting the angles of the CKM Unitarity Triangle,
providing constraints on the CKM parameters. 
  
The penguin transitions that are expected to be present in many of the 
rare hadronic  B decays might be a source of new physics as new particles,
not predicted by the Standard Model,  can be present in the loop transitions.
Insights into the new physics can be learned from these rare decays.

\vspace{0.2cm}
The field of rare B decays  has expended rapidly with the operation of the new $B$ 
factories, the 
asymmetric-energy $e^+e^-$ colliders PEP-II at SLAC and KEKB at KEK. Each one 
produced more than 100million $B\bar{B}$ pairs allowing the BABAR and Belle 
experiments 
to make the observation of new decay channels or precise measurements of
 channels previously  observed, mostly by the CLEO experiment.

Due to space restrictions,
this review covers only a limited number of $B_u$ and $B_d$ (herein referred to as
$B$) decay channels into mesonic final states, namely:
\mbox{$B \ra \pi \pi, K \pi, KK$}, \mbox{$B \ra (\eta \eta\prime)(K,K^{*},\rho,\pi)$},
\mbox{$B \ra \rho K, \rho \pi$} and \mbox{$B \ra \phi K^{*}, \rho \rho, \rho K^*$}.

The experimental values of BF and $A_{CP}$, as measured by the BABAR, Belle and
CLEO experiments at the time of this conference (July 2004),
as well as the average values calculated by the Heavy Flavor Averaging Group (HFAG)
\cite{HFAG} are presented.

\vspace{-0.2cm}
\section{\mbox{$B \ra \pi \pi, K \pi, KK$} decays}

\begin{table*}[htb]
\caption{BF ($\times 10^{-6}$) of the $\pi \pi$, $K \pi$ and $KK$ decay modes.} 
\label{table:1}
\newcommand{\cc}[1]{\multicolumn{1}{c}{#1}}
\renewcommand{\tabcolsep}{0.7pc} 
\renewcommand{\arraystretch}{1} 
\begin{tabular}{@{}llllllll}

\hline
Mode    	& BABAR & Ref.&  Belle & Ref. & CLEO & Ref. & Average        \\
\hline
$\pippim$       & \mbox{$4.7\pm 0.6\pm 0.2$}            & \cite{b11}   
                & \mbox{$4.4\pm 0.6\pm 0.3$}            & \cite{b21}
                & \mbox{$4.5^{+1.4+0.5}_{-1.2-0.4}$}    & \cite{b41}
                & \mbox{$4.6\pm 0.4$}                   \\
$\pippiz$	& \mbox{$ 5.5^{+1.0}_{-0.9}\pm 0.6$}    & \cite{b19}    
                & \mbox{$5.0\pm 1.2\pm 0.5$}            & \cite{b21} 
                & \mbox{$4.6^{+1.8+0.6}_{-1.6-0.7}$}    & \cite{b41} 
                & \mbox{$5.2\pm 0.8$}                   \\
$\pizpiz$	& \mbox{$2.1\pm 0.6\pm 0.3$}            &  \cite{b13}
                & \mbox{$ 1.7\pm 0.6\pm 0.2$}           & \cite{b25} 
                & \mbox{$<4.4$}                         & \cite{b41} 
                & \mbox{$1.9\pm 0.5$}                   \\
\hline
$\kppim$	& \mbox{$17.9\pm 0.9\pm 0.7$}           & \cite{b11}  
                & \mbox{$18.5\pm 1.0\pm 0.7$}           & \cite{b21} 
		& \mbox{$18.0^{+2.3+1.2}_{-2.1-0.9}$}   & \cite{b41} 
                & \mbox{$18.2\pm 0.8$}                  \\
$\kppiz$	& \mbox{$12.8^{+1.2}_{-1.1}\pm 1.0$}    & \cite{b19}  
                & \mbox{$12.0\pm 1.3^{+1.3}_{-0.9}$}    & \cite{b21} 
                & \mbox{$12.90^{+2.4+1.2}_{-2.2-1.1}$}  & \cite{b41} 
                & \mbox{$12.5^{+1.1}_{-1.0}$}           \\
$\kzpip$	& \mbox{$22.3\pm 1.7\pm 1.1$}           & \cite{b1}  
                & \mbox{$22.0\pm 1.9\pm 1.1$}           & \cite{b21} 
                & \mbox{$18.8^{+3.7+2.1}_{-3.3-1.8}$}   & \cite{b41} 
                & \mbox{$21.8\pm 1.4$}                  \\
$\kzpiz$	& \mbox{$11.4\pm 1.7\pm 0.8$}           & \cite{b1} 
                & \mbox{$11.7\pm 2.3^{+1.2}_{-1.3}$}    & \cite{b21} 
                & \mbox{$12.8^{+4.0+1.7}_{-3.3-1.4}$}   & \cite{b41}  
                & \mbox{$11.7\pm 1.4$}                  \\
\hline
$\kpkm$	        & \mbox{$<0.6$}	& \cite{b11} 
                & \mbox{$<0.7$} & \cite{b21} 
                & \mbox{$<0.8$} & \cite{b41} 
                & \mbox{$<0.6$} \\
$\kpkzb$        & \mbox{$<2.5$}	& \cite{b1} 
                & \mbox{$<3.3$} & \cite{b21} 
                & \mbox{$<0.7$} & \cite{b41} 
                & \mbox{$<0.7$} \\
$\kzkzb$        & \mbox{$<1.8$}	& \cite{b1} 
                & \mbox{$<1.5$} & \cite{b21} 
                & \mbox{$<3.3$} & \cite{b41} 
                & \mbox{$<1.5$} \\
\hline
\end{tabular}\\
\end{table*}

\begin{table*}[htb]
\caption{$A_{CP}$ of the $\pi \pi$, $K \pi$ and $KK$ decay modes.}
\label{table:2}
\newcommand{\cc}[1]{\multicolumn{1}{c}{#1}}
\renewcommand{\tabcolsep}{0.1pc} 
\renewcommand{\arraystretch}{1} 
\begin{tabular}{@{}llllllll}

\hline
Mode    	& BABAR & Ref. &  Belle & Ref. & CLEO & Ref. & Average        \\

\hline
$\pippim$       &                                               &  
                &  $+0.53 \pm 0.19 $                            & \cite{ex0401} 
                &                                               &
                & $+0.53 \pm 0.19 $                             \\
$\pippiz$	& \mbox{$-0.03^{+0.18}_{-0.17}\pm 0.02$}        & \cite{b19} 
                & \mbox{$+0.00\pm 0.10\pm 0.02$}                & \cite{a20} 
                &                                               &
                & \mbox{$-0.01\pm 0.09$}                        \\
\hline
$\kppim$	& \mbox{$-0.107\pm 0.041\pm 0.013$}             & \cite{a4} 
                & \mbox{$-0.088\pm 0.035\pm 0.018$}             & \cite{a4} 
		& \mbox{$-0.04\pm 0.16$}                        & \cite{a15} 
                & \mbox{$-0.095\pm 0.028$}                      \\
$\kppiz$	& \mbox{$-0.09\pm 0.09 \pm 0.01$}               & \cite{b19} 
                & \mbox{$+0.06\pm 0.06\pm 0.02$}                & \cite{a20} 
                & \mbox{$-0.29\pm 0.23\pm 0.02$}                & \cite{a15} 
                & \mbox{$-0.00\pm0.05$}                         \\
$\kzpip$	& \mbox{$-0.05\pm 0.08 \pm 0.01$}               & \cite{b1} 
                & \mbox{$+0.05\pm 0.05\pm 0.01$}                & \cite{b4} 
                & \mbox{$+0.18\pm 0.24\pm 0.02$}                & \cite{a15} 
                & \mbox{$+0.03\pm0.04$}                         \\
$\kzpiz$	& \mbox{$+0.03\pm 0.36\pm 0.11$}                & \cite{b1} 
                & \mbox{$+0.16\pm 0.29\pm 0.05$}                & \cite{a20} 
                &                                               &
                & \mbox{$+0.11\pm 0.23$}                        \\
\hline
\end{tabular}\\
\end{table*}

\begin{table*}[htb]
\caption{BF ($ \times 10^{-6}$) of the ($\eta, \eta \prime)(K,K^*,\rho, \pi$) 
decay modes.} 
  
\label{table:3}
\newcommand{\cc}[1]{\multicolumn{1}{c}{#1}}
\renewcommand{\tabcolsep}{0.8pc} 
\renewcommand{\arraystretch}{1} 
\begin{tabular}{@{}llllllll}

\hline
Mode    	& BABAR &  Ref. & Belle & Ref. & CLEO & Ref. & Average        \\
\hline
$\eta\prime K^+$    & \mbox{$76.9\pm 3.5\pm 4.4$}            & \cite{b3} 
                    & \mbox{$78\pm 6\pm 9$}                  & \cite{b22} 
		    & \mbox{$80^{+10}_{-0.9}\pm 7$}          & \cite{b42} 
                    & \mbox{$77.6^{+4.6}_{-4.5}$}            \\
$\eta K^+$          & \mbox{$3.4\pm 0.8\pm 0.2$}             & \cite{b16} 
                    & \mbox{$5.3^{+1.8}_{-1.5}\pm 0.6$}      & \cite{b24} 
                    & \mbox{$2.2^{+2.8}_{-2.2}$}             & \cite{b42} 
                    & \mbox{$3.7\pm 0.7$}      \\
$\eta \prime K^0$   & \mbox{$60.6\pm 5.6\pm 4.6$}            & \cite{b3} 
                    & \mbox{$68\pm 10^{+9}_{-8}$}            & \cite{b22} 
                    & \mbox{$89^{+18}_{-16}\pm 9$}           & \cite{b42} 
                    & \mbox{$65.2^{+6.0}_{-5.9}$}            \\
$\eta K^0$          & \mbox{$<5.2$}                          & \cite{b16} 
                    &                                        &  
                    & \mbox{$<9.3$}                          & \cite{b42} 
                    & \mbox{$<5.2$}                          \\
\hline
$\eta\prime K^{*+}$    & \mbox{$<14$}                        & \cite{b4} 
                       & \mbox{$<90$}                         & \cite{b22} 
	               & \mbox{$<35$}                        & \cite{b42} 
                       & \mbox{$<14$}                        \\
$\eta K^{*+}$          & \mbox{$25.6\pm 4.0\pm 2.5$}         & \cite{b4} 
                       & \mbox{$26.5^{+7.8}_{-7.0}\pm 3.0$}  & \cite{b22} 
                       & \mbox{$26.4^{+9.6}_{-8.2}\pm 3.3$}  & \cite{b42} 
                       & \mbox{$25.9^{+3.8}_{-3.6}$}         \\
$\eta \prime K^{*0}$   & \mbox{$<7.6$}                       & \cite{b4} 
                       & \mbox{$<20$}                        & \cite{b22} 
                       & \mbox{$<24$}                        & \cite{b42} 
                       & \mbox{$<7.6$}                       \\
$\eta K^{*0}$          & \mbox{$18.6\pm 2.3 \pm1.2$}         & \cite{b4} 
                       & \mbox{$16.5^{+4.6}_{-4.2}\pm 1.2$}  & \cite{b22} 
                       & \mbox{$13.8^{+5.5}_{-4.6}\pm 1.6$}  & \cite{b42} 
                       & \mbox{$17.5^{+2.2}_{-2.1}$}         \\
\hline

$\eta  \rho^+$         & \mbox{$<14$}                        & \cite{b4} 
                       & \mbox{$<6.2$}                       & \cite{b22} 
		       & \mbox{$<15$}                        & \cite{b42} 
                       & \mbox{$<6.2$}                       \\
$\eta \rho^0$          & \mbox{$<1.5$}                       & \cite{b4} 
                       & \mbox{$<5.5$}                       & \cite{b24} 
                       & \mbox{$<10$}                        & \cite{b42} 
                       & \mbox{$<1.5$}                       \\
$\eta \pi^+$      & \mbox{$5.3\pm 1.0 \pm 0.3$}              & \cite{b16} 
                  & \mbox{$5.7^{+1.4}_{-1.7}\pm 0.9$}        & \cite{b24} 
                  & \mbox{$1.2^{+2.8}_{-1.2}$}               & \cite{b42} 
                  & \mbox{$4.9^{+0.9}_{-0.8}$}               \\
$\eta \pi^0$      & \mbox{$<2.5$}                            & \cite{b4}
                  &                                          &
                  & \mbox{$<2.9$}                            & \cite{b42} 
                  & \mbox{$<2.5$}                            \\

\hline
$\eta \prime \rho^+$        & \mbox{$<22$}                   & \cite{b4}
                            &                                &
	                    & \mbox{$<33$}                   & \cite{b42} 
                            & \mbox{$<22$}                   \\
$\eta \prime \rho^{0}$      & \mbox{$<4.3$}                  & \cite{b4} 
                            & \mbox{$<14$}                   & \cite{b22}
                            & \mbox{$<12$}                   & \cite{b42} 
                            & \mbox{$<4.3$}                  \\
$\eta \prime \pi^{+}$       & \mbox{$<4.5$}                  & \cite{b16} 
                            & \mbox{$<7$}                    & \cite{b23}
                            & \mbox{$<12$}                   & \cite{b42} 
                            & \mbox{$<4.5$}                  \\
$\eta \prime \pi^{0}$       & \mbox{$<3.7$}                  & \cite{b4} 
                            &                                &
                            & \mbox{$<5.7$}                  & \cite{b42} 
                            & \mbox{$<3.7$}                  \\
\hline
\end{tabular}\\
\end{table*}

\begin{table*}[htb]
\caption{$A_{CP}$ of the ($\eta, \eta \prime)(K,K^*,\rho, \pi$) decay modes.} 
\label{table:5}
\center
\newcommand{\cc}[1]{\multicolumn{1}{c}{#1}}
\renewcommand{\tabcolsep}{3pc} 
\renewcommand{\arraystretch}{1} 
\begin{tabular}{@{}lll}

\hline
Mode    	& BABAR    &         Ref.       \\

\hline
$\eta \prime K^+$  & \mbox{$+0.04\pm 0.05\pm 0.01$}     & \cite{b3}  \\                 
$\eta K^+$	   & \mbox{$-0.52\pm 0.24 \pm 0.01$}    & \cite{b16} \\
\hline
$\eta K^{*+}$      & \mbox{$+0.13\pm 0.14\pm 0.02$}     & \cite{b4}  \\
$\eta K^{*0}$      & \mbox{$-0.02\pm +0.11\pm 0.02$}    & \cite{b4}  \\
\hline
$\eta \pi^+$       & \mbox{$-0.44\pm 0.18\pm 0.01$}     & \cite{b16} \\              
$\eta \rho^+$	   & \mbox{$+0.06\pm 0.29 \pm 0.02$}    & \cite{b4}  \\                 \hline     
\end{tabular}\\
\end{table*}

\hspace{0.5cm}Two-body decays of $B$ to combinations of pions and/or kaons provide 
information that can be used to determine the angles $\alpha$ and $\gamma$
($\phi_2$ and $\phi_3$ in another common  notation) of the CKM Unitarity Triangle.
Precise measurements of their BFs allow us to set limits
on the theoretical hadronic uncertainties that affect 
the extraction of these 
angles.

The decay amplitudes of these processes have contributions from tree and/or 
penguin amplitudes.
The contribution of the penguin amplitude can be determined from BF
measurements and the direct CP violation can be established with $A_{CP}$
measurements. 

\vspace{0.2cm}
The current BF measurements for these two-body decays are summarised in Table 
\ref{table:1}.

$B \ra \pi^0 \pi^0$ decay mode was  
recently observed by the BABAR and Belle experiments with a statistical significance 
of 4.2 and 3.4 standard deviations, respectively. The measured BF´s
are higher than the theoretical predictions of less than $10^{-6}$ \cite{NP606}. 

The BFs of $K^+\pi^{-,0}$ decay modes were found by all three
experiments to be much higher than the BFs of $\pi^+\pi^{-,0}$ decay
modes. As the former modes are dominated by penguin transitions, this experimental
result demonstrates that the  penguin amplitude is significant.
 
The measured BFs of $\pi^+ \pi^-$ and $\pi^+ \pi^0$ decay
modes indicate a ratio $2 \Gamma (\pi^+ \pi^0)/ \Gamma (\pi^+ \pi^-) \simeq 2$,
where $\Gamma$ represents the decay rate of $B$ into the specified final state.
The theoretical calculations that assume the dominance of the tree amplitudes
in both decays \cite{PRD65} predict a ratio equal to unity. This discrepancy
indicates a significant penguin contribution to the $\pi^+ \pi^-$ decay amplitude, 
the so called ``penguin pollution''. 

$B \ra KK$ decays are modes where the rescattering effects are expected to be
significant. The available data only set upper 
limits on their BFs. Their comparison with perturbative 
QCD predictions \cite{PRD63} does not yet indicate evidence for rescattering 
effects. 

\vspace{0.2cm}
The current $A_{CP}$ measurements of the specified two-body decays are presented 
in Table \ref{table:2}.

Significant direct CP violation was reported by the Belle collaboration  for the 
$B \ra \pi^+ \pi^-$ decay mode. The report was  based on a time-dependent
analysis \cite{ex0401} that is not discussed here. A preliminary time-integrated
CP asymmetry measurement was also reported: \mbox{$+0.53 \pm 0.19$}. 

No other significant direct CP asymmetry was reported for these decays.
 It is worth observing, however,
that both BABAR and Belle experiments measured a non-zero asymmetry
in the $K^+\pi^-$ decay mode with a statistical significance higher than two 
standard deviations.

\section{$B \ra (\eta, \eta \prime)(K,K^*,\rho, \pi$) decays}

\begin{table*}[htb]
\caption{BF ( $\times 10^{-6}$) of the $\rho \pi$ and $\rho K$ decay modes.}
\label{table:6}
\newcommand{\cc}[1]{\multicolumn{1}{c}{#1}}
\renewcommand{\tabcolsep}{0.65pc} 
\renewcommand{\arraystretch}{1} 
\begin{tabular}{@{}llllllll}

\hline
Mode    	& BABAR & Ref. & Belle & Ref. & CLEO & Ref.& Average        \\

\hline

$\rho^+ \pi^-$     & \mbox{$22.6\pm 1.8 \pm 2.2$}             & \cite{b7} 
                   & \mbox{$29.1^{+5.0}_{-4.9}\pm 4.0$}       & \cite{b30} 
                   & \mbox{$27.6^{+8.4}_{-7.4}\pm 4.2$}       & \cite{b44} 
                   & \mbox{$24.0\pm 2.5$}                     \\ 
$\rho^+ \pi^0$     & \mbox{$10.9\pm 1.9 \pm 1.9$}             & \cite{b9} 
                   & \mbox{$13.2\pm 2.3^{+1.4}_{-1.9}$}       & \cite{b37} 
                   & \mbox{$<43$}                             & \cite{b44} 
                   & \mbox{$12.0\pm 1.9$}                     \\ 
$\rho^0 \pi^+$     & \mbox{$9.5\pm 1.1 \pm 0.8$}              & \cite{b9} 
                   & \mbox{$8.0^{+2.3}_{-2.0}\pm 0.7$}        & \cite{b32} 
                   & \mbox{$10.4^{+3.3}_{-3.4}\pm 2.1$}       & \cite{b44} 
                   & \mbox{$9.2^{+1.2}_{-1.1}$}               \\ 
$\rho^0 \pi^0$     & \mbox{$<2.9$}                            & \cite{b9} 
                   & \mbox{$5.1\pm 1.6 \pm 0.9$}              & \cite{ex0401} 
                   & \mbox{$<5.5$}                            & \cite{b44}
                   & \mbox{$5.1\pm 1.6 \pm 0.9$}              \\
\hline
$\rho^- K^+$       & \mbox{$7.3^{+1.3}_{-1.2}\pm 1.3$}        & \cite{b7} 
                   & \mbox{$15.1^{+3.4+2.4}_{-3.3-2.6}$}      & \cite{b35}    
                   & \mbox{$16^{+8}_{-6}\pm 3$}               & \cite{b44} 
                   & \mbox{$9.0\pm 1.6$}                      \\ 
$\rho^0 K^+$       & \mbox{$3.9\pm 1.2 ^{+1.3}_{-3.5}$}       & \cite{b10} 
                   & \mbox{$3.9\pm 0.6^{+0.8}_{-0.4}$}        & \cite{b36} 
                   & \mbox{$8.4^{+4.0}_{-3.4}\pm 1.8$}        & \cite{b44} 
                   & \mbox{$4.1^{+0.9}_{-0.7}$}               \\ 
\hline     
\end{tabular}\\
\end{table*}

\begin{table*}[htb]
\caption{$A_{CP}$ of the $\rho \pi$ and $\rho K$ decay modes.}
\label{table:7}
\newcommand{\cc}[1]{\multicolumn{1}{c}{#1}}
\renewcommand{\tabcolsep}{1.25pc} 
\renewcommand{\arraystretch}{1} 
\begin{tabular}{@{}llllll}

\hline
Mode    	& BABAR & Ref. &  Belle & Ref. & Average        \\

\hline

$\rho^+ \pi^-$  & \mbox{$-0.11\pm 0.06\pm 0.03$}             & \cite{a4}
                & \mbox{$-0.38^{+0.19+0.04}_{-0.21-0.05}$}   & \cite{b35}
		& \mbox{$-0.14\pm 0.06$}                     \\
$\rho^+ \pi^0$  & \mbox{$+0.24\pm 0.16\pm 0.06$}             & \cite{a13}
                & \mbox{$+0.06\pm 0.19+0.04$}                & \cite{a34}
		& \mbox{$0.16\pm 0.13$}                      \\
$\rho^0 \pi^+$  & \mbox{$-0.19\pm 0.11\pm 0.02$}             & \cite{a13}
                &                                            &
	        &  \mbox{$-0.19\pm 0.11$}                    \\
\hline
$\rho^- K^+$	& \mbox{$+0.18\pm 0.12\pm 0.08$}             & \cite{a4}
                & \mbox{$+0.22^{+0.22+0.06}_{-0.23-0.02}$}   & \cite{b35}
                &  \mbox{$+0.19 \pm 0.12$}                   \\

\hline     
\end{tabular}\\
\end{table*}

\hspace{0.5cm}$B$ decays to final states containing an $\eta$ or an $\eta \prime$
provide useful information for understanding the relative contribution
of the tree and penguin amplitudes and have potential for establishing the
direct CP violation in the Standard Model.

\vspace{0.2cm}
The decay amplitudes of \mbox{$B \ra (\eta, \eta \prime)(K^+,K^{*+})$} 
have contributions from a CKM suppressed tree $b \ra u $ amplitude and 
from  two $b \ra s$ penguin amplitudes that suffer a destructive interference 
\cite{PLB254}.
An additional contribution of a flavor-singlet penguin amplitude  to 
 the $\eta \prime K^{*+}$ decay amplitude is expected \cite{0308}.
The interference between the two penguin amplitudes and the $\eta/\eta \prime$ mixing 
angle make the $\eta \prime K^+$ and $\eta K^{*+}$ decay modes to be 
significantly enhanced relative to 
$\eta K^+$ and $\eta \prime K^{*+}$ decay modes, respectively. A similar mechanism is 
expected 
for $K^0$ and $K^{*0}$ decay modes except there is no tree amplitude contribution.

The experimental BF measurements, summarized in Table \ref{table:3}, confirm 
the predicted relationship between the BFs of these decays. Their absolute values 
are in agreement with recent next-to-leading-order QCD calculations \cite{NPB651}.

\mbox{$B \ra (\eta,\eta \prime)(\rho,\pi)$} decay amplitudes are dominated 
by an external  tree $b \ra u$ amplitude while an internal tree $b \ra u$ 
amplitude is color 
suppressed and  two penguin $b \ra s$ amplitudes are CKM suppressed. 
The interference between the two penguin amplitudes is reduced while the interference 
between the tree and penguin ones is significant, enhancing the $\eta^{(\prime)}\rho^+$ 
decay modes.

The present amount of experimental data provided only upper limits 
 for the BFs of most of these decay modes. They are presented in Table \ref{table:3}.
It can be observed that the upper limit of $\eta \prime \rho^+$ BF is much higher 
than of the other modes.

\vspace{0.2cm}
The destructive interference between the two penguin amplitudes of the 
$\eta ^{(\prime ^)}K^{(*)}$ decay modes make the 
interference between the CKM suppressed tree amplitude and the penguin amplitudes  
significant for the suppressed modes, leading to large $A_{CP}$. 
Large $A_{CP}$ are expected for $\eta \pi$ decay mode as well.

Preliminary $A_{CP}$ measurements are reported by the BABAR collaboration for
six decay modes. They are shown in Table \ref{table:5}. The asymmetry for 
$\eta \prime K^*$ decay mode was also measured by the Belle 
($-0.02 \pm 0.07 \pm 0.01$ \cite{a17}) and CLEO
($+0.03 \pm 0.12 \pm 0.02$ \cite{a15}) experiments.

While no significant CP asymmetry can be concluded from these measurements,
the results for $\eta K^+$ and $\eta \pi^+$ can be noted as being non-zero
with a statistical significance higher than two standard deviations.

\section{$B \ra \rho \pi,\rho K$ decays}

\begin{table*}[htb]
\caption{BF ($\times 10^{-6}$) of the $\phi K^*$, $\rho \rho$
and $\rho K^*$ decay modes.}
\label{table:8}
\newcommand{\cc}[1]{\multicolumn{1}{c}{#1}}
\renewcommand{\tabcolsep}{0.8pc} 
\renewcommand{\arraystretch}{1} 
\begin{tabular}{@{}llllllll}

\hline
Mode    	& BABAR & Ref. &  Belle & Ref. & CLEO & Ref. & Average        \\

\hline
$\Phi K^{*+}$     & \mbox{$12.7^{+2.2}_{-2.0}\pm 1.1$}      & \cite{b15} 
                  & \mbox{$6.7^{+2.1+0.7}_{-1.9-1.0}$}      & \cite{b34} 
                  & \mbox{$10.6^{+6.4+1.8}_{-4.9-1.6}$}     & \cite{b43} 
                  & \mbox{$9.7\pm 1.5$}                     \\ 
$\Phi K^{*0}$     & \mbox{$11.2\pm 1.3\pm 0.8$}             & \cite{b15} 
                  & \mbox{$10.0^{+1.6+0.7}_{-1.5-0.8}$}     & \cite{b34} 
                  & \mbox{$11.5^{+4.5+1.8}_{-3.7-1.7}$}     & \cite{b43} 
                  & \mbox{$10.7\pm 1.1$}                    \\ 
\hline
$K^{*+} \rho^0$     & \mbox{$10.6^{+3.0}_{-2.6} \pm 2.4$}   & \cite{b15} 
                    &                                       &
                    & \mbox{$<74$}                          & \cite{b45} 
                    & \mbox{$10.6^{+3.8}_{-3.5}$}           \\                  
\hline
$\rho^+ \rho^-$     & \mbox{$30\pm 4 \pm 5$}                & \cite{b20} 
                    &                                       &
                    &                                       &  
                    & \mbox{$30 \pm 6$}                     \\ 
$\rho^+ \rho^0$     & \mbox{$22.5^{+5.7}_{-5.4} \pm 5.8$}   & \cite{b15} 
                    & \mbox{$31.7 \pm 7.1^{+3.8}_{-6.7}$}   & \cite{b31}
                    &                                       &
                    & \mbox{$26.4^{+6.1}_{-6.4}$}           \\ 
$\rho^0 \rho^0$     & \mbox{$<2.1$}                         & \cite{b15} 
                    &                                       &
                    & \mbox{$<18$}                          & \cite{b45} 
                    & \mbox{$<2.1$}                         \\ 
\hline     
\end{tabular}\\
\end{table*}

\begin{table*}[htb]
\caption{$A_{CP}$ of the $\phi K^*$, $\rho \rho$, $\rho K^*$ decay modes.}
\label{table:9}
\newcommand{\cc}[1]{\multicolumn{1.2}{c}{#1}}
\renewcommand{\tabcolsep}{1.2pc} 
\renewcommand{\arraystretch}{1} 
\begin{tabular}{@{}llllll}

\hline
Mode    	& BABAR & Ref. &  Belle & Ref. & Average        \\

\hline
$\Phi K^{*+}$     & \mbox{$+0.16\pm 0.17\pm 0.03$}          & \cite{b15}
                  & \mbox{$-0.13\pm 0.29^{+0.08}_{-0.11}$} & \cite{a29}  
                  & \mbox{$+0.15\pm 0.15$}                  \\ 
$\Phi K^{*0}$     & \mbox{$+0.04\pm 0.12\pm 0.02$}          & \cite{b15}
                  & \mbox{$+0.07\pm 0.15^{+0.05}_{-0.03}$}  & \cite{a29}  
                  & \mbox{$+0.05\pm 0.10$}                  \\ 
\hline
$K^{*+} \rho^0$   & \mbox{$+0.20^{+0.32}_{-0.29} \pm 0.04$} & \cite{b15}
                  &                                        &   
                  &  \mbox{$+0.20^{+0.32}_{-0.29}$}         \\                  
\hline
$\rho^+ \rho^0$   & \mbox{$-0.19\pm 0.23 \pm 0.03$}        & \cite{b15}
                  & \mbox{$+0.00 \pm 0.22 \pm 0.03$}        & \cite{b18}  
                  &  \mbox{$-0.09\pm 0.16$}                \\  
\hline     
\end{tabular}\\
\end{table*}

\hspace{0.5cm}$B$ decaying to $\rho \pi$ and $\rho K$ final states are important decay 
channels for constraining the CKM parameters and for testing the theoretical models
that predict the BFs in a quite large range.

The isospin analysis of the $\rho \pi$ decay amplitudes is one of the main method for
determining the angle $\alpha$  of the CKM Unitarity Triangle.
The decay amplitudes  are extracted from the BF and $A_{CP}$ experimental values.

\vspace{0.2cm}
The current BF measurements of these decays are shown in Table \ref{table:6}.

The $\rho^0 \pi^0$ decay mode, very much needed for the above mentioned isospin analysis,
was recently measured by the Belle experiment with a statistical significance of 3.5
standard deviations. The measured BF is higher than most of the theoretical 
predictions, suggesting that some contributions 
to the decay amplitude might be higher than expected, making the isospin analysis more 
complicated.

\vspace{0.2cm}
$A_{CP}$ was measured for four decay modes and the results are shown in Table
\ref{table:7}. No asymmetry is observed in these decays.

\section{$B \ra \phi K^*, \rho \rho, \rho K^*$ decays }

\hspace{0.5cm}$B$ decays to vector-vector final states are of particular interest as more
observables are available for CP violation tests: the angular distributions and 
the vector meson polarization components (as well as BF and $A_{CP}$).
In addition, $\rho \rho$ decay modes, being very similar to $\rho \pi$ modes, 
provide an alternative method for extracting the angle $\alpha$ of the 
CKM Unitarity Triangle. 

\vspace{0.2cm}
$B \ra \Phi K^*$ decay proceeds through a pure $b \ra s$ penguin 
transition offering the possibility of an unambiguous signature for penguins.
As the penguin transitions are the same for $\Phi K^{*0}$ and $\Phi K^{*+}$ 
decay modes, the difference being only the spectator quark, similar BFs 
are expected. Indeed, the measurements performed by the three 
experiments, summarized in Table \ref{table:8}, confirm these expectations. 
	
$B \ra \rho \rho$ modes proceed through $b \ra u$ tree and CKM suppressed $b \ra d$ 
penguin transitions. Their measured BFs are also shown in Table \ref{table:8}.
While both BABAR and Belle experiments observed the $\rho^+ \rho^0$ decay mode,
BABAR collaboration also reported the observation of the $\rho^+ \rho^-$ decay
mode with a statistical significance of 5.1 standard deviations and lowered the
limit on $\rho^0 \rho^0$ BF.

\begin{table*}[htb]
\caption{The longitudinal polarization fraction
of the $\phi K^*$, $\rho \rho$, $\rho K^*$ decay modes.}
\label{table:10}
\center
\newcommand{\cc}[1]{\multicolumn{1}{c}{#1}}
\renewcommand{\tabcolsep}{3.5pc} 
\renewcommand{\arraystretch}{1} 
\begin{tabular}{@{}lllll}

\hline
Mode    	& BABAR & Ref.       \\

\hline
$\Phi K^{*+}$     & \mbox{$0.46\pm 0.12\pm 0.03$}      & \cite{f3} \\
$\Phi K^{*0}$     & \mbox{$0.52\pm 0.07\pm 0.02$}      & \cite{f4} \\
\hline
$K^{*+} \rho^0$   & \mbox{$0.96^{+0.04}_{-0.15} \pm 0.04$}   & \cite{f3} \\  
\hline
$\rho^+ \rho^-$   & \mbox{$0.99\pm 0.03^{0.04}_{-0.03}$}      & \cite{f3} \\
$\rho^+ \rho^0$   & \mbox{$0.97^{+0.03}_{-0.07} \pm 0.04$}    & \cite{f3} \\  
\hline     
\end{tabular}
\end{table*}

\vspace{0.2cm}
The $A_{CP}$ is expected to be large for $\rho K^*$ decay modes that proceed through
both $b \ra u$ tree and $b\ra s$ penguin transitions. This asymmetry is expected 
to be small in the Standard Model and sizable in the presence of new physics effects
in the other decay channels.

The current $A_{CP}$ measurements, summarized in Table \ref{table:9}, do not
indicate significant CP asymmetry in these decays.

\vspace{0.2cm}
In the framework of the perturbative QCD \cite{PRD66}, the longitudinal polarization 
fraction, $f_L=\frac{\Gamma_L}{\Gamma}$, is predicted to be very close to 100\% for all 
 vector-vector decay modes.
This prediction is only partially confirmed by the present measurements.
The values measured by the BABAR experiment, presented in Table \ref{table:10},
indicate $f_L$ close to 100\% for the $\rho K^*$
and $\rho \rho$ decay modes and around 50\% for the $\phi K^*$ decay modes.
Similar results were obtained by the Belle experiment for the $\rho^+ \rho^0$
mode (\mbox{$0.948\pm 0.106 \pm 0.021$ \cite{f2}}) and the $\phi K^{*0}$ mode 
(\mbox{$0.43 \pm 0.19 \pm 0.04$ \cite{f1}}). Different theoretical interpretations 
exist for this result, some of them suggesting new physics effects in the $b \ra s$ 
penguin transition \cite{ph031}.

\section{Conclusions}

\hspace{0.5cm}The already impressive amount of data accumulated by the new B factories at SLAC and KEK
yielded important advances in the field of rare hadronic B decays. A few of them are
summarized here.

Precise measurements of BF and $A_{CP}$ were performed for many decay channels
and new decay modes were observed. Two of these new modes are of particular importance: \mbox{$B \ra \pi^+ \pi^-$} with a BF of \mbox{$(1.9 \pm 0.5)\cdot
10^{-6}$} and \mbox{$B \ra \rho^0 \pi^0$} with a BF of \mbox{$(5.1 \pm 1.6 \pm 0.9) \cdot 10^{-6}$}.

The observation of direct CP violation in the $B \ra \pi^+ \pi^-$ decay mode,
in a time-dependent analysis, confirmed through a preliminary time-integrated
CP asymmetry measurement of $+0.53 \pm 0.19$, was reported by the Belle collaboration.

Non-zero direct CP asymmetries were measured for $K^+\pi^-$, $\eta K^+$ and $\eta \pi^+$
decay modes with statistical significances higher than two standard deviations.
Updates of these measurements are eagerly expected.

The longitudinal polarization fraction of vector-vector decay modes was experimentally
found close to 100\% for $\rho \rho$ and $\rho K^*$ modes, confirming the theoretical
predictions, and around 50\% for $\phi K^*$ modes, much lower than the theoretical
predictions. Theoretical efforts are needed in order to understand this discrepancy.

\end{document}